\documentclass[preprint,aps]{revtex4}

\usepackage{graphicx}
\usepackage{dcolumn}
\usepackage{bm}

\begin{document}


\title{High Field Phenomena of Qubits}

\author{J. van Tol}
\email{vantol@magnet.fsu.edu}
\altaffiliation{Corresponding author, Dr. Johan van Tol, National High Magnetic Field
Laboratory, Florida State University, 1800 E. Paul Dirac Dr, Tallahassee, FL 32310,
USA. Tel (+1) 850 644 1187}
\affiliation{Center for Interdisciplinary Magnetic Resonance, National High Magnetic Field
Laboratory, Florida State University, Tallahassee, Florida-32310,
USA}
\author{G.~W. Morley}
\affiliation{London Center for nanotechnology and Department of Physics and Astronomy, University College London,
London, WC1H 0AH, UK}
\author{S. Takahashi}
\affiliation{Department of Physics and Center for Terahertz Science and Technology, University of California, Santa Barbara, California 93106}
\author{D.~R. McCamey}
\author{C. Boehme}
\affiliation{Department of Physics, University of Utah, Salt Lake City, Utah 84112, USA}
\author{M.~E. Zvanut}
\affiliation{Department of Physics, University of Alabama at Birmingham, Birmingham, Alabama 35294-1170, USA}

\date{\today}

\begin{abstract}
Electron and nuclear spins are very promising candidates to serve as quantum bits (qubits) for proposed quantum computers, as the spin degrees of freedom are relatively isolated from their surroundings, and can be coherently manipulated e.g. through pulsed EPR and NMR. For solid state spin systems, impurities in crystals based on carbon and silicon in various forms have been suggested as qubits, and very long relaxation rates have been observed in such systems. We have investigated a variety of these systems at high magnetic fields in our multi-frequency pulsed EPR/ENDOR spectrometer. 
A high magnetic field leads to large electron spin polarizations at helium temperatures giving rise to various phenomena that are of interest with respect to quantum computing. For example, it allows the initialization of the both the electron spin as well as hyperfine-coupled nuclear spins in a well defined state by combining millimeter and RF radiation ; it can increase the T$_2$ relaxation times by eliminating decoherence due to dipolar interaction; and it can lead to new mechanisms for the coherent electrical readout of electron spins.  We will show some examples of these and other effects in Si:P, SiC:N, and nitrogen-related centers in diamond.
\end{abstract}

\maketitle

\section{Introduction}

Electron and nuclear spins have been recognized as particularly interesting for quantum computing applications. The coupling between the spin and orbital degrees of freedom can be very small, giving rise to relatively long relaxation times. Moreover, coherent manipulation of these spins had been utilized in both Electron Paramagnetic Resonance (EPR) and Nuclear Magnetic Resonance (NMR) as a powerful spectroscopic technique for years before the birth of quantum computing. High Frequency EPR and ENDOR (Electron Nuclear Double Resonance) can be used advantageously to characterize qubit systems. The main focus of this paper, though, is to illustrate that the properties of qubits can depend on the magnetic field. In particular we will focus on the advantages that high magnetic fields can provide in the initialization, manipulation, and the read-out of qubit systems. 

A large variety of spin systems have been proposed for quantum computing applications, while quantum computation has been demonstrated using liquid state NMR using a 7-spin quantum computer~\cite{vandersypen01}. Liquid state NMR, however, suffers  from limitations related to the relatively long time necessary to perform a quantum manipulation ($>1$ ms), due to the weak coupling between nuclear spins. Also scalability to a larger number of qubits is difficult if not impossible~\cite{Warren97}. 

Electron spins, or combinations of electron spins and nuclear spins, seem to be a more promising candidate for qubits or qubit ensembles in the long term. It has been shown that entanglement between electron and nuclear spins can be achieved~\cite{Mehring03, Mehring04} and that combinations of microwave pulses and RF pulses, addressing electron and nuclear spin transitions respectively, can form a basis for quantum computation~\cite{Rahimi04, Sato07, Scherer08}. A number of solid state systems have shown that very long relaxation times can be reached. In this context a figure of merit has been defined which as the number of operations that can be performed on the qubit before coherence is lost~\cite{Yannoni99}, and is usually defined as $Q_M=T_2/T_{\mathrm{op}}$. It is obvious that good qubits must have a sufficiently long coherence time or $T_2$. This time can be increased by high magnetic fields.

The requirements for the spin-lattice relaxation time ($T_1$) are somewhat more ambiguous. Obviously, the $T_1$ must be long enough to allow a long $T_2$  However, in some of the proposed qubit systems the electronic spin-lattice relaxation time is several orders of magnitude longer than the spin-spin relaxation or spin-memory relaxation time. It is not so obvious whether that is an advantage. For example, in pulsed EPR this means that the shot-repetition time can get very long, of the order of seconds or minutes thereby increasing total measurement times and limiting sensitivity. For quantum computers it may mean a very long reset time to a well-defined initial state, as the timescale involved will be of the order of $T_1$. To some degree is might be possible to tune the $T_1$ with the temperature. However, the magnetic field can also play a role to shorten the $T_1$.

Finally, one of the great challenges for quantum computing is the read-out of the result of the quantum computation. An obvious readout mechanism is to use conventional EPR or NMR. In such experiments, the properties of the system are measured as a macroscopic magnetic moment that is the result of an ensemble of microscopic spins that are oscillating in phase at a particular moment in time. The typical numbers of spins that is necessary to form a measurable macroscopic oscillating magnetic moment in a single pulse is $\approx 10^9$ for electron spins and $\approx 10^{15}$ for nuclear spins. This is the approach used for the liquid state NMR quantum computation and while such an ensemble quantum computer has some advantages in quantum error correction, it is not likely to be the approach that will lead to success.  Below, we will discuss another approach, which utilizes EPR as a control mechanism and electrical readout, where single spin sensitivity has been demonstrated \cite{Xiao04}.

\section{Materials and Methods}

Almost all the results discussed here are performed with the multi frequency superheterodyne quasi-optical spectrometer at the National High Magnetic Field Laboratory in Tallahassee~\cite{Vantol05, Morley08}. The main operating frequencies of the spectrometer are 120, 240, and 336 GHz. EPR measurements at 9.7 GHz were performed on a Bruker Elexsys 580 spectrometer.

The phosphorus doped silicon (Si:P) for the nuclear polarization experiments was a $3 \times 3 \times 1$ mm piece of crystalline silicon from Wacker Siltronic with [P]= $1 \times 10^{15}$/cm$^{-3}$. The silicon sample used in the electrically detected magnetic resonance was a 0.33 mm thick (111) oriented prime grade Cz-grown crystalline silicon wafer with [P]$\approx 10^{15}$ cm$^{-3}$~\cite{McCamey08}. The
type-Ib diamond (Sumitomo electric industries) had a density of N impurities of 10$^{19}$ to 10$^{20}$ cm$^{-3}$. The $1 \times 1 \times 1$ mm sample was irradiated with 1.7 MeV electrons with a dose of $5 \times 10^{17}$ cm$^{-3}$ and subsequently annealed at 900 C for 2 hours. The 4H-SiC $4 \times 4 \times 1$ mm single crystal was grown by the physical vapor transport (PVT) method with [N]$\approx 7 \times 10^{16}$ cm$^{-3}$.

\begin{figure}
\includegraphics{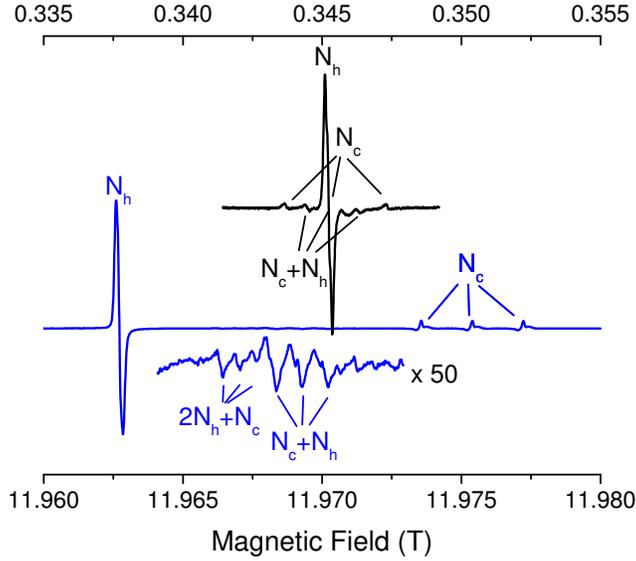}
\caption{CW-EPR spectra of nitrogen dopants in 4H-SiC at 9.7 and 336 GHz at 10 K and 20 K respectively. The high resolution at 336 GHz confirms the presence of N$_c$--N$_h$ exchange coupled pairs, giving a signal at $ g = (g_{Nh} + g_{Nc})/2$, while a further triplet at $ g = (2g_{Nh} + g_{Nc})/3$ is tentatively ascribed to a exchange coupled center formed by two N$_h$ sites and one N$_c$ site. Note that the N$_c$ center intensity is reduced due to partial saturation.}
\label{figSiC}
\end{figure}

\begin{figure}
\includegraphics{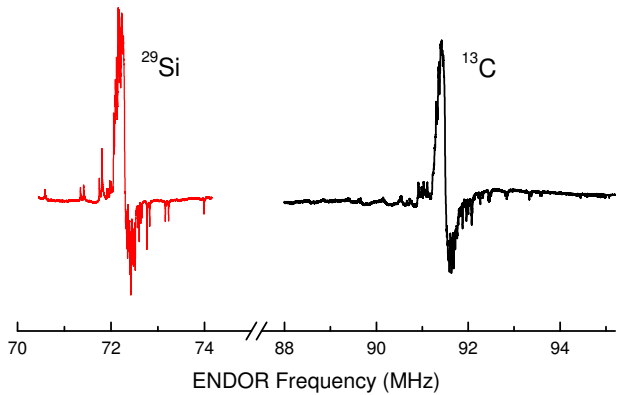}
\caption{Pulsed Mims ENDOR spectra of the N$_h$ center in 4H-SiC at 240 GHz.}
\label{figSiCENDOR}
\end{figure}

\section{Results and Discussion}

\subsection{Resolution at high fields}
Spectral resolution can be an important factor for qubit systems, as it is desirable to be able to selectively address individual qubits. It is e.g. possible to apply gate-voltages in order to shift the resonance position through changes in g-values and/or hyperfine splitting~\cite{Kane98}. Also, systems with alternating types of qubits have been proposed~\cite{Lloyd93, Benjamin02}. While at high fields the hyperfine resolution in single crystals is field independent, the g-value resolution increases linearly with the operating frequency. An example in given in figure~\ref{figSiC}, showing spectra of nitrogen centers in 4H-SiC at 9.7 GHz and 336 GHz with $B \parallel c$. The signals of hexagonal nitrogen centers (N$_h$), cubic nitrogen centers (N$_c$), and pair centers~\cite{Zvanut07} have a strong overlap at 9.7 GHz, while they are well separated at 336 GHz. One can therefore, at least in theory, envision chains of alternating N$_h$ and N$_c$ defects, with the possibility of selective excitation. Especially for selective operations performed by pulsed EPR with an excitation bandwidth $\Delta \omega$ given by $\Delta \omega = 2 \pi / t_p$ with $t_p$ the pulse-length~\cite{SchweigerBook} a large separation is necessary. Additionally, if nuclear (ENDOR) transitions of coupled nuclei are to be addressed, high fields enable separation of transitions of different nuclei through differences in nuclear Zeeman splitting. An example for the N$_h$ centers in 4H-SiC is shown in the pulsed ENDOR spectrum in Figure~\ref{figSiC}b. At 240 GHz (8.55 T) the transitions from the $^{29}$Si nuclei and those of the $^{13}$C are well separated, whereas they have considerable overlap at X-band frequencies.

\subsection{Frequency dependence of spin-lattice relaxation}

\begin{figure}
\includegraphics{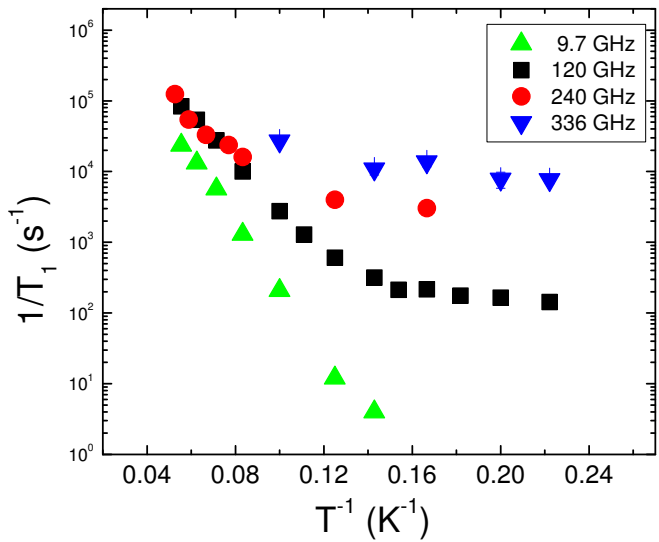}
\caption{Spin relaxation rate $T_1^{-1}$ for the hexagonal nitrogen center in 4H-SiC as a function of temperature at various frequencies. The error margins are of the order of the symbol size.}
\label{figT1}
\end{figure}

At the lowest temperatures the spin lattice relaxation (SLR) tends to be dominated by the direct process, and we have investigated a number of slow-relaxing qubit systems with spin-lattice relaxation at different frequencies at low temperatures. The low-temperature SLR process of Cr$^{5+}$ spins in K$_3$NbO$_8$  is close to 3 order of magnitude faster at 240 GHz than it is at X-band frequencies~\cite{Nellutla08}. Another example is shown in Figure~\ref{figT1}  for the
hexagonal nitrogen center (N$_h$) in silicon carbide, where measured spin lattice relaxation rates are shown for 9.7 GHz, 120, GHz, 240 GHz, and 336 GHz, in the temperature range of 4-20 K. At higher temperatures the rate seems to be determined by an Orbach process that involves thermal excitation to a level roughly 50 cm$^{-1}$ above the ground state. At low temperatures the relaxation rate is proportional to the temperature, indicating a direct spin-lattice relaxation process. At X-band the $T_1$ becomes of the order of seconds and hard to measure at temperatures below 6 K, whilst at 336 GHz the direct process limits the $T_1$ to about 100 $\mu$s, more than 4 orders of magnitude faster. The values of $T_1$ at 120, 240, and 336 GHz in the low-temperature range display a $\omega^4$ dependence as opposed to the $\omega^2$ dependence found for the Cr$^{5+}$ system~\cite{Nellutla08}.

\subsection{Quenching of spin-spin interactions}

The spin-memory time or spin-spin relaxation time is related to incoherent changes in the local field of the electron spin. These can be changes in the local crystal field e.g. the hopping from one Jahn-Teller minimum to another, but more often these are related to changes in the local magnetic field induced by fluctuations of the surrounding  nuclear or electron spins. In typical low-concentration organic radicals the hyperfine coupling to surrounding protons usually limits the low-temperature $T_2$ to 1-2 $\mu$s, and deuteration can significantly increase $T_2$.  In systems with little or no super-hyperfine interaction the $T_2$ can be quite long. For example for phosphorus impurities in natural abundance silicon (4.69 \% $^{29}$Si), the hyperfine coupling with the $^{29}$Si limits the T$_2$ to a few hundred microseconds, while in isotopically pure $^{28}$Si the $T_2$ can be extended to the millisecond range~\cite{Tyryshkin03}. In those cases, the (dipolar) electron-electron spin-spin interactions will tend to limit the $T_2$, unless the concentration is extremely low.  

The incoherent changes in the local field of the observer spins (decoherence) can be caused by neighboring spins changing their spin state due to $T_1$ type processes, or via $T_2$ type processes. The latter is usually the dominant factor, and can be approximated by a spin flip-flop process,  in which two spins in different spin states exchange their spin state without a net energy change.  For a $S=1/2$ system the probability of this process will be proportional to the product of the spin-up and spin-down populations:  $P_{\mbox{flip-flop}} \sim  1/(e^{(h \nu /2kT)} + e^{-(h \nu / 2kT)})^2 = (2 \cosh(h \nu / 2 kT))^{-2}$~\cite{Kutter95}. This implies that by going to the limit of $h \nu >> kT$ this process can be quenched to a large degree, as we recently showed for the nitrogen and N-V (nitrogen-vacancy) centers in synthetic diamond~\cite{Takahashi08}.   

In the case of these diluted centers in diamond, the $T_2$ relaxation times are already quite long at room temperature ($\approx 6 \mu$s), and increase in the limit $kT \ll h\nu$ to several hundreds of $\mu $s, and are most likely limited by $^{13}$C hyperfine interactions. However, in more concentrated spin systems like single crystals of molecular magnets, the dipolar electron-electron interactions have prevented even a direct measurement of the $T_2$ relaxation, as it tends to be too fast. By utilizing  high frequencies in combination with low temperatures it will become possible to study spin dynamics in these kind of systems by quenching the decoherence induced by the dipolar coupled electron spin bath. While pulsed spin resonance has so far been limited to dilute spin systems, as are encountered in biological systems and lightly doped diamagnetic compounds, these developments might lead to new applications in more concentrated spin systems as often found in solid state physics.

\subsection{Initialization - Large Electronic and Nuclear Spin Polarization}

The equilibrium population of the spin sublevels is given by the Boltzmann distribution, and for a S=1/2 system the relative population difference or spin polarization P ($P = (p_\downarrow - p_\uparrow )/(p_\downarrow + p_\uparrow) $) corresponds to $\tanh(h\nu /2kT)$. As a frequency of 240 GHz corresponds to 11.43 K, a 99.99\% electron spin polarization is reached at 2.1 K. It is thus possible to study by EPR the effects of such a large spin polarization (sometimes referred to as the saturated paramagnetic phase) at more or less standard $^4$He temperatures. One such effect is spin-spin relaxation which was discussed in the previous section. 

\begin{figure}
\includegraphics{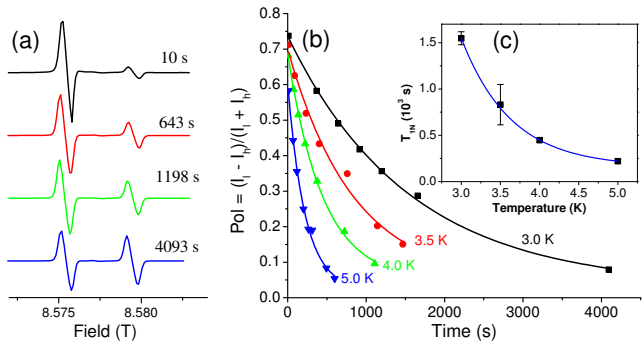}
\caption[Nuclear Polarization of $^{31}$P donor sites in silicon.]{Nuclear Polarization of $^{31}$P donor sites in silicon. (a) EPR spectra at 3 K taken at various intervals after irradiating the high-field transition for 5 minutes. (b) $^{31}$P nuclear polarization as a function of time at various temperatures. (c) Nuclear spin-lattice relaxation $T_{1N}$ as a function of temperature, with a mono-exponential fit.}
\label{SiPfigure}
\end{figure}

For electron spin qubits this creates the possibility of simple initialization of the system in a pure electronic spin state. Furthermore, if hyperfine interactions are also important, like in the silicon-based quantum computer proposed by Kane~\cite{Kane98}, the nuclear spins can also be polarized to a very high degree. One way is simply to use the Overhauser effect. For example, in the previously mentioned silicon based system with phosphorus impurities, the $^{31}$P hyperfine interaction is purely isotropic. Therefore the hyperfine interaction will slightly allow transitions that involve flip-flops of electron and nuclear spins, as $ a {\vec S} \cdot {\vec I} $ can be written as $a(S_z \cdot I_z + \frac{1}{2}(S^+I^- + S^-I^+))$. On the other hand, flip-flip transitions induced by terms of the type $S^+I^+$ and $S^-I^-$ remain strictly forbidden.  In the case of phosphorus in silicon, this means that by saturating the high-field hyperfine transition at high fields and low temperatures, the population will end up in the other hyperfine state. This is what has been done for the data shown in figure~\ref{SiPfigure}a. Here the EPR spectrum of phosphorus in silicon ($\approx 10^{15}$/cm$^3$) at 3 K is measured at very low power after saturating the high-field hyperfine component for 5 minutes at high power. While initially a polarization of the order of 75\% can be reached, this polarization is found to decay back to equilibrium on a timescale of half an hour, which simply corresponds to the nuclear spin-lattice relaxation time $T_{1N}$. At 5 K, this relaxation time is reduced to about 3.5 minutes. In this (limited) temperature range, $T_{1N}$ can be fitted with a exponential as $T_{1N}^{-1} = e^{-\Delta E/kT}$ with $\Delta E = 14 \pm 2$ K. This latter energy is close to the electron Zeeman splitting of 11.5 K, and we conclude that the nuclear spin-lattice relaxation is limited by thermal excitation to the upper-electron spin level through the very same partially allowed electron-nuclear flip-flop transition.

This Overhauser process does not allow nuclear polarization in the other direction (i.e. anti-polarization or negative spin-temperature). Saturating the low-field hyperfine transition does not increase the intensity of the high-field line. However by applying radio-frequency (RF) waves resonant with the ENDOR transition at the same time in cw mode~\cite{Morley07, Brill02} or sequentially in pulsed mode~\cite{Morley08b}, the polarization can be achieved in both directions, while the polarization process itself is much more efficient and reduces the time needed by several orders of magnitude. 

\subsection{Readout}

\begin{figure}
\includegraphics{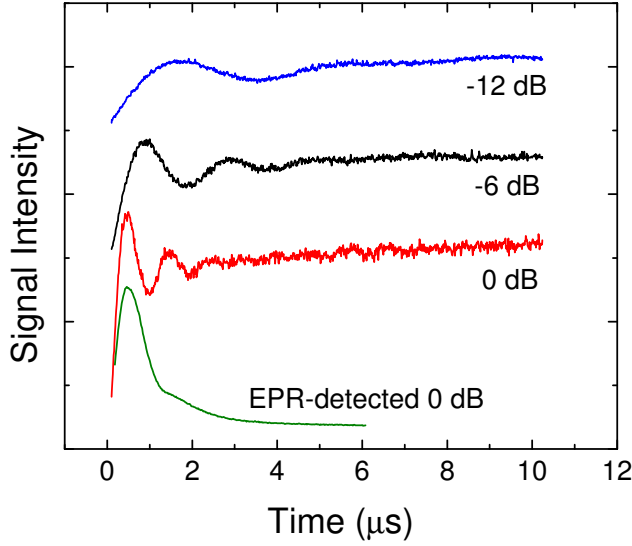}
\caption{EDMR and EPR-detected Rabi oscillations in P:Si at 240 GHz. The upper three traces show the integrated electrical response as a function of the pulse length at 0, 6, and 12 dB of attenuation. The lower trace shows the EPR-detected Rabi-oscillation by detecting the echo height as a function of length of the first pulse in a two-pulse Hahn-echo sequence. }
\label{figRabi} 
\end{figure}

While single spin detection cannot be achieved via conventional EPR or NMR detection techniques, both optical and electrical detection schemes have achieved single spin detection in some systems related to quantum computing~\cite{Wrachtrup06, Elzerman04, Xiao04}. An important question is to what extent the detection scheme itself contributes to decoherence, and until recently the maximum coherence time measured via electrical detection was limited to 2 microseconds~\cite{Huebl08}. However, recently we have shown that for electrical detection of phosphorus spins in silicon at high fields, long coherence times, of the order of 100 $\mu$s are preserved at measurements at 8.5 T~\cite{Morley08c}. Figure~\ref{figRabi} shows Rabi-oscillations measured at 240 GHz in a sample of silicon with gold contacts for electrically detected magnetic resonance (EDMR) detection~\cite{McCamey08}. 

The sample is mounted in a Fabry-Perot resonator, but the silicon thickness is 350 $\mu$m, which is close to one wavelength at 240 GHz, taking the high index of refraction into account. The sample is also larger than the beam-waist in the resonator, which means that we have a large distribution of $B_1$ fields inside the sample, leading to very strong damping of the Rabi oscillations when measured in EPR detection, as shown in Figure~\ref{figRabi}. In electrical detection, the surface is much smaller (~0.1 mm$^2$), while only the centers at a limited depth from the contacts are sampled. Indeed well-defined Rabi-oscillations are observed. The Rabi damping rate ( $ \approx 10 \mu $s) is still much faster than the decoherence rate~\cite{Morley08c} which we ascribe to the remaining $B_1$ inhomogeneity due to the contacts and the finite depth of the current through the sample. By going to isotopically pure $^{28}$Si even longer times might be obtained. 

While the experiments so far focused on the mechanisms of the spin-dependent conduction, for actual devices the sensitivity will be an issue.  For these samples, the surface was still relatively large (~0.1 mm$^2$) and the single shot sensitivity has been estimated at around 10$^{-7}$ spins~\cite{Morley08c}. The effect on the current was quite significant and for some conditions $\Delta I/I$ exceeded 10\%. This, in view of the fact that with a $B_1$ field with a maximum value of 0.3 Gauss only one of the two hyperfine components is only partially excited, which corresponds to a very significant effect on the current. As the signal-to-noise ratio does not necessarily depend on sample size~\cite{McCamey08}, the prospect of single spin detection by limiting the active area does not seem out of reach~\cite{McCamey06}.

\section{Summary}
High frequencies and high magnetic fields can play a crucial role in the study and perhaps implementations of qubit systems for quantum computing. The magnetic field can have a strong influence on both $T_1$ and $T_2$ relaxation times, initialization in a pure quantum state can be easily achieved at high fields and low temperatures, and read-out via electrical (charge) detection at high fields is very promising.

The high-frequency pulsed spectrometer used in this research is available for experiments by outside users in the context of the National High Magnetic Field Laboratory user program (see http://users.magnet.fsu.edu). 

\subsection{Acknowledgements}
This work was supported in part by the NSF Grant
No. DMR-0520481. The
National High Magnetic Field Laboratory is supported by
NSF Cooperative Agreement No. DMR-065411, by the
state of Florida, and by the DOE.

\bibliography{cairns_final_arxiv}

\end{document}